\documentstyle[11pt,newpasp,twoside,epsf]{article}
\def\edcomment#1{\iffalse\marginpar{\raggedright\sl#1\/}\else\relax\fi}
\reversemarginpar

\begin{document}
\title{The MAGIC Cherenkov Telescope for gamma ray astronomy}
 \author{Josep Flix, for the MAGIC Collaboration (*)}
\affil{Institut de F\'{\i}sica d'Altes Energies, Edifici Cn, Universitat Autonoma de
Barcelona, Cerdanyola del Vall\'{e}s, 08193, Spain \\
(*) Updated collaborator list at: http://www.magic.iac.es
}

\begin{abstract} The Major Atmospheric Gamma ray Imaging Cherenkov Telescope (MAGIC) is
in commissioning phase and will start to become fully operative by
the end of 2003. Located at {\it El Roque de los Muchachos} in La
Palma (Canary Islands, Spain), it has the largest reflector area
(17 m diameter) of all the existing Cherenkov telescopes. New
technologies have been used to reduce the energy threshold for
gamma-ray detection to about 30 GeV. Due to its characteristics,
the catalog of very high energy sources will considerably increase
with the MAGIC observations, anticipating exciting results for the
near future. An overview of the telescope, its current status and
first results, together with a highlight of the scientific
research is presented.
\end{abstract}

\section{Introduction}

Gamma ray astronomy has provided in the las few years spectacular
results, led by the success of the CGRO ({\it Compton Gamma Ray
Observatory}) satellite mission which revealed that the high
energy Universe was more exciting than expected. During its
lifetime, the EGRET telescope onboard the CGRO provided an all-sky
survey above 100 MeV consisting of 271 sources (Hartman et al.
1999). Apart of the remarkable detection of 66 AGN Blazars, the
vast majority of these sources are still unidentified. Almost
during the same period, the ground-based gamma ray astronomy has
also been developed. The technique consists of the detection of
the atmospheric Cherenkov light emitted by the particle showers
initiated by gamma radiation on entering into the atmosphere. This
Cherenkov flash lasts for a few {\it ns}. The Whipple Cherenkov
telescope opened this ground-based era by the detection of the
Crab Nebulae in 1989. Then, extragalactic Blazar sources not seen
by satellites were also discovered by using this technique. Other
instruments based on the same principle have confirmed the Whipple
results (HEGRA, CANGAROO, CAT) and nowadays about a dozen gamma
ray sources have been detected to emit in the TeV energy range.
The observations done by satellites measured well below around 10
GeV, while the existing Cherenkov telescopes have detected sources
above 300 GeV. This energy gap, which is still virtually
unexplored, is really important to understand which is the origin
of the cut-off on the spectra which will explain the lack of
sources measured by the Cherenkov telescopes in comparison to
EGRET.

The MAGIC (Major Atmospheric Gamma Imaging Cherenkov) telescope
was designed in 1998 (Barrio et al. 1998) with the main goal of
being the Imaging Atmospheric Cherenkov Telescope (IACT) with the
lowest gamma energy threshold possible with the technological
improvements affordable and based on the experience acquired with
the first generation of Cherenkov telescopes. By using this
detection technique, which provides much large effective areas
(and much superior flux sensitivity) than satellite detectors,
good angular resolution, acceptable energy resolution and a well
tested capability to separate gammas from backgrounds, eventually
a plethora of new sources will be discovered since for most of
known sources the energy spectrum is of power-law nature and
therefore they should exhibit a much higher flux in that energy
region than at higher energies.

\begin{figure}[h]
\plotfiddle{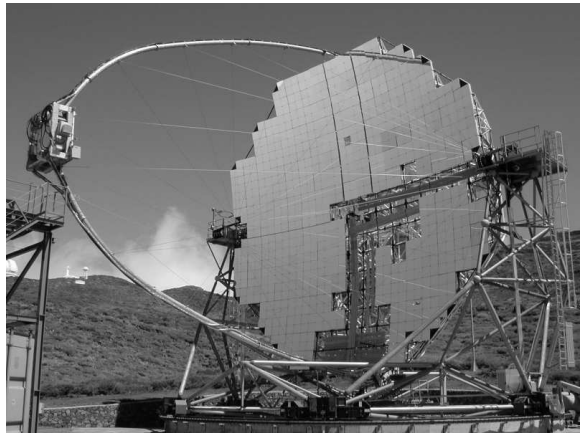}{5,5cm}{0}{27}{27}{-115.}{0.}
\caption{View of the MAGIC Telescope in August 2003. Almost all the mirrors are already on place.}
\end{figure}

The 17 m diameter f/D=1 MAGIC telescope is the largest of the new
generation of IACTs. MAGIC is located in the Canarian island of La
Palma (28.8 N, 17.9 W) at the {\it Roque de los Muchachos}
observatory (ORM), 2200 m above sea level. Its 241 m$^2$ parabolic
dish is composed of 964 $49.5\times 49.5$ cm$^2$ all-aluminum
spherical mirror tiles mounted on a lightweight ($< 10$ ton)
carbon fiber frame. The parabolic shape was chosen to minimize the
time spread of the Cherenkov light flashes on the camera plane,
which allows to reduce the rate of fake events induced by
night-sky background light. Mirrors are grouped in panels of three
or four, which can be oriented during the telescope operation
through a novel active mirror control system to correct for the
possible deformations of the telescope structure.

The camera is made of 577 good quantum efficiency, fast
photomultipliers with hemispherical photocatode that allows for
light double-crossing. Each photomultiplier is coupled to a small
light collecting cone to maximize the active surface of the
camera. An special wavelength-shifting coating provides red
extended sensitivity and allows for light-trapping, which
increases the photomultiplier effective quantum efficiency. The
total field of view of the camera is of about 4$\deg$. The
photomultiplier signals are transmitted to a distant Control House
($\sim$150 m) by using analog optical fiber signals. Signals are
processed by a multilevel trigger system and 300 MHz FADC are used
for pulse digitalization.

While all other new generation Cherenkov telescopes aim at the improvement of sensitivity and
energy resolution in the 100 GeV regime by using stereoscopic systems of relatively small
(10 m) telescopes, MAGIC, through the choice of a single, larger reflector, will achieve the
lowest energy threshold among IACTs, of about 30 GeV. Its altazimuth mount can point to anywhere
in the sky in about 20 seconds, a unique feature which is essential for the study of
transient events like GRBs.

\section{The Physics Goals}

The main research targets that will be addressed by MAGIC are of a
wide nature. They cover subjects as:

\begin{itemize}

\item Measurements of the AGNs energy flux above 30 GeV which will allow
to determine the gamma ray horizon and eventually extract the cosmological parameters
and the extragalactic background light density. Fast flares will be used to constrain
Quantum Gravity effects. Multi-wavelength observation campaigns will allow to determine
the emission processes that occurs in these sources.

\item The systematic study of galactic gamma emitters such as Supernova Remnants,
Plerions, X-ray binaries, Pulsars, unidentified EGRET sources, etc.
where the observation in this energy range might allow to discriminate between
different acceleration mechanisms and might hopefully lead to the
identification of the main sources of cosmic rays up to about 10$^{15}$ GeV.

\item Due to the MAGIC fast slewing, the observation of Gamma Ray Bursts in
the new energy window will be possible.

\item Due to its low energy threshold and high sensitivity, MAGIC will be a good
instrument for searches of Dark Matter annihilation signals into gamma rays
(Flix et al. 2003).

\end{itemize}

\section{The Present Status}

The construction of the foundation for the MAGIC telescope started in
September 2001 and just few months later the whole telescope structure was
completed (December 2001). Mirror installation started in summer 2002 and now of
about 200 m$^2$ of mirror area are already on place (Figure 1). The Active Mirror Control has
proven to be very precise and fast. The telescope drive system was installed
during past year and has recently being commissioned up the highest speed
which turns out to lead to a maximum repositioning time of 20 seconds. The tracking
system has been calibrated using bright star pointings. The camera was installed on the
site in November 2002 and has been commissioned in March 2003 after the winter break.

The 1st and 2nd level trigger systems have been already installed
and commissioned as well as the whole computing system for the
telescope control and DAQ. The installation of the FADC system was
completed after some on-place tests and measurements with the
whole readout chain have already been performed. Almost all
subsystems are integrated into a central control. During June 2003
first Cherenkov images were recorded by using the whole Data
Acquisition Chain (Figure 2), as well as light pulses using a
novel calibration system, which is being used to characterize the
whole telescope.

The construction of a definitive Control House with an unique shape inspired upon
the island Dancing-Dwarf traditions has almost reached completion. All the
systems are being moved the the Control House, before the official inauguration
which will take place on the 10th October 2003. The telescope system are being extensively and
intensively checked now and observations of the Crab Nebulae (the standard candle for high energy sources)
by this winter will allow us to understand and check the whole detector. We expect to
start regular observations by the end of this year 2003.

\begin{figure}[t]
\plotfiddle{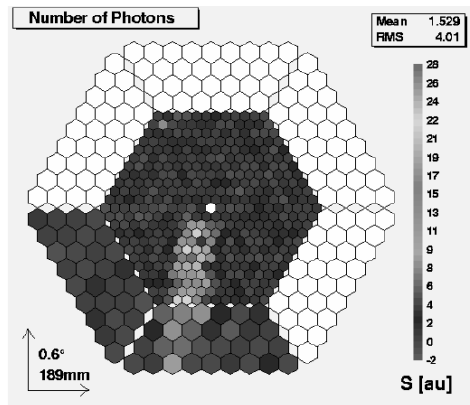}{5,4cm}{0}{34}{34}{-95.}{0.}
\caption{One of the first air showers registered with MAGIC. Not all the outer pixels of the camera
were connected to the readout.}
\end{figure}

\section{summary}

The MAGIC telescope is in its final commissioning phase and it is expected to start
regular observations by the end of this year. If the telescope behaves as expected,
it will soon be able to provide exciting results on a wide variety of astrophysical phenomena.

\begin{acknowledgements}
  J. Flix would like to acknowledge all the organizers for the very nice and
  productive conference.
\end{acknowledgements}

\end{document}